\documentclass[12pt]{article}

\usepackage{amssymb}

\def\fnote#1#2{\begingroup\def\thefootnote{#1}\footnote{#2}\addtocounter{footnote}{-1}\endgroup}

\begin{document}

\hfill \phantom{ {\bf Submission date:} \today}

\vskip .8truein

\centerline{\large {\bf Black Hole Probes of Automorphic Space}}

\vskip .4truein

\centerline{{\sc Rolf Schimmrigk}\fnote{1,}{email contacts: netahu@yahoo.com; rschimmr@iusb.edu}\fnote{2}{On leave 
        of absence from Indiana University South Bend, USA}}

\vskip .2truein
 \parskip .1truein
 \baselineskip=17pt

 \centerline{Theory Group}

 \centerline{Department of Physics, CERN}

 \centerline{CH-1211, Geneva 23, Switzerland}

\vskip 1truein

\baselineskip=18pt

 \centerline{\bf Abstract}

 Over the past few years the arithmetic Langlands program has proven useful in addressing physical 
problems.  In this paper it is shown how Langlands' reciprocity conjecture for automorphic forms, 
in combination with a representation theoretic notion of motives, suggests a framework in which the 
entropy of automorphic black holes can be viewed as a probe of spacetime that is sensitive to the 
geometry of the extra dimensions predicted by string theory.
 If it were possible to produce black holes with automorphic entropy in the laboratory their evaporation 
 would provide us with information about the precise shape of the compact geometry.

\vskip .5truein

\centerline{\it This essay received an honorable mention in the 2012 Essay Competition }
 
\centerline{\it of the  Gravity Research Foundation}

\vskip .4truein


\renewcommand\thepage{}
\newpage
\parindent=0pt

\baselineskip=20.5pt
\parskip=.1truein

\renewcommand\thepage{}
\newpage
\parindent=0pt

 \pagenumbering{arabic}

\section{Introduction}\label{ra_sec1}

Black holes have played an essential role in developing our understanding of fundamental 
aspects of gravity. In particular their role as thermodynamical objects has been one of the 
central issues ever since the paradigmatic insights of Bekenstein and Hawking. More 
recently, the structure of black holes with higher supersymmetry has been illuminated in 
fundamentally new ways by the application of tools from the theory of modular forms of 
higher rank. The emergence of
such automorphic forms makes it possible to link the theory of
supersymmetric black holes to the arithmetic Langlands program \cite{rl79},
thereby making it possible to introduce ideas from the theory of
automorphic representations as new tools that can be used to
elucidate the implications of the entropy results obtained so far.

One of the key questions raised by the Langlands program is whether
general algebraic automorphic forms are geometric in the sense
that there exist geometric structures, called motives, whose
properties encode exactly the same information as the automorphic
form. The roots of this problem reach back into the late 19th century, but only its
 modern incarnation as the framework of automorphic motives is far-reaching enough 
to have found an application in string theory, 
in particular the problem of an emergent spacetime \cite{rs08}. The emergence of
automorphic black hole entropy functions makes it therefore  
natural to ask whether they encode motivic information
about the compact part of spacetime. This leads to a new point of view of
(automorphic) black holes as physical probes that are
sensitive to the nature of the extra dimensions. The automorphic nature of
the entropy implies that if one were to experiment with such black
holes in the laboratory, a finite number of measurements would
suffice to completely determine these functions. This point of view is quite 
general, depending mostly on the notion of duality, but is developed 
here in the context of a special class of black holes for the sake of simplicity.

\section{Automorphic black holes as probes of extra dimensions}

Over the past decade impressive progress has been made toward
the resolution of a problem that is almost 40 years old - the
microscopic understanding of the entropy of black holes. It has
proven useful to focus on black holes with extended
supersymmetries because such objects are simple enough,
but not too simple. More recently, it was shown in particular for certain types
of black holes in ${\cal N}=4$ supersymmetric theories that their
entropy is encoded in the Fourier coefficients of
Siegel modular forms, automorphic forms that provide some of the
simplest generalizations of classical modular forms of one
variable with respect to congruence subgroups of the full modular
group ${\rm SL}(2,{\mathbb Z})$.

The general conceptual framework of automorphic entropy functions
has not been formalized yet, but a rough ouline that encodes the
existing concrete examples can be formulated as follows. Suppose we
have a theory which contains scalar fields parametrized by a
homogeneous space $\prod_k (G_k/H_k)$ of Lie groups $G_k$ 
and subgroups $H_k$, leading to electric and
magnetic charge vectors $Q=(Q_e,Q_m)$ in a lattice
$\Lambda$. Assume further that the theory has a discrete T-duality group 
$\prod_k D_k({\mathbb Z})$ defined over the rational integers ${\mathbb Z}$ and 
that the charge vector $Q$ leads to norms $||Q||_i, ~i=1,...,r$ which are 
T-duality invariant. Choosing variables $\tau_i, i=1,...,r$ conjugate to the
invariant charge norms $||Q||_i$ one can consider  automorphic forms
$\Phi(\tau_i)$ defined on the resulting generalized upper half plane ${\cal H}_r$ 
spanned by the $\tau_i$. 

The idea is that by endowing the charge norms with an appropriate integral
structure ${\mathbb Z} \ni k_i \sim ||Q||_i$, the Fourier expansion of such
 automorphic forms 
 \begin{equation}
  \Phi(\tau_i) ~=~ \sum_{k_n} g(k_1,...,k_r)
                    q_{1}^{k_{1}} \cdots q_{r}^{k_{r}} , ~~~~~q_k := e^{2\pi i \tau_k},
 \end{equation}
  determines the automorphic entropy of the black holes.
 Writing the expansion of the partition function $Z(\tau_i)$ as 
  \begin{equation}
  Z(\tau_i) ~=~ \frac{1}{{\widetilde \Phi}(\tau_i)}
     ~=~ \sum_{k_n} d(k_1,...,k_r) q_{1}^{k_{1}} \cdots q_{r}^{k_{r}},
 \end{equation}
 leads to the microscopic entropy as a function of the charges as
 \begin{equation}
  S_{\rm mic}(Q) ~\sim ~ \ln d(||Q||_1,...,||Q||_r).
 \end{equation}
 Here ${\widetilde \Phi}$ denotes a slight modification of the Siegel form
 $\Phi$ that is determined by the vanishing behavior of $\Phi$.

The above outline encompasses the behavior of the entropy of 
black holes in certain ${\cal N}=4$ compactifications obtained by considering ${\mathbb
Z}_N-$quotients of the heterotic toroidal compactification ${\rm Het}(T^6)$ (or their 
type IIA duals),  a class of 
models first considered in ref. \cite{chl95}. Specifically, it was shown in \cite{dvv96,js05,gk09}
 that for these CHL$_N$ models the microscopic entropy of extreme
 Reissner-Nordstr\o m type black holes is described by genus two Siegel
 modular forms $\Phi^N(\tau_1,\tau_2,\tau_3) \in S_w(\Gamma^{(2)}_0(N))$,
 where the weight $w$ is determined by the order $N$ of the quotient
 group. Here the automorphic groups are Hecke type congruence subgroups
 of the symplectic group, $\Gamma_0^{(2)}(N) \subset {\rm Sp}(4,{\mathbb Z})$.

\section{From Siegel entropy to automorphic geometry}

Given black holes whose entropy is determined by automorphic
forms one can ask whether the compactification manifolds leads to motives rich 
enough to support these automorphic forms. While it is not expected that general
automorphic forms are of motivic origin,  the subset of forms of algebraic type are 
conjectured to have a geometric origin \cite{y01, p11}.

In the special case of genus two Siegel forms the conjectures
concerning their motivic origin indicate that the compactification
manifold cannot provide the appropriate motivic cycle
structure if one follows the picture developed in arithmetic geometry. 
The quickest way to see this is by noting that the Hodge decomposition of a
pure spinor motive $M_\Phi$ associated to a genus two Siegel form $\Phi$ takes the form
 \begin{equation}
  H(M_\Phi) ~=~ H^{2w-3,0} \oplus H^{w-1,w-2} \oplus
                H^{w-2,w-1} \oplus H^{0,2w-3}.
 \label{siegel-hodge}\end{equation}
For mixed motives this relation can change even for classical forms \cite{kls10}. 
While the Hodge type (\ref{siegel-hodge}) is that 
of a Calabi-Yau variety, the precise structure is only correct for modular forms of weight
  three,  which is too restrictive for the CHL$_N$  Siegel modular forms whose weights 
 can be as high as $w=10$. Hence for most CHL$_N$ models the Siegel modular forms
cannot be induced directly by motives in the way usually envisioned in mathematics. 

In the face of this first obstruction it is useful to recognize that the Siegel modular forms 
which appear in the context of CHL$_N$ black hole entropy are not of
general type. Instead, they are obtained by combining the Skoruppa lift \cite{nps92} from
classical modular forms to Jacobi forms with the Maa\ss~lift \cite{m79} from
Jacobi forms to Siegel modular forms
 \begin{equation}
  f_{w+2}(\tau) \in S_{w+2} ~\stackrel{\rm SL}{\longrightarrow} ~\varphi_{w,1}(\tau,\rho) \in J_w
                      ~\stackrel{\rm ML}{\longrightarrow} ~\Phi_w(\tau,\sigma,\rho) \in S_w.
 \label{ms-lift}\end{equation}
While the Maa\ss-Skoruppa lift represents an important step, it does not immediately solve the problem because 
the motivic support $M_f$ for classical modular forms $f$ of weight $w$ leads to the Hodge decomposition
 \begin{equation}
 H(M_f) ~=~ H^{w-1,0} \oplus H^{0,w-1},
 \end{equation}
 hence the only modular forms that can fit into heterotic
 compactifications have weight two, three, or four.
 Both of the above obstructions to a geometric interpretation of black hole automorphic 
forms can be overcome, as described in the remainder of this paper.
 
The key to the identification of the motivic origin of the CHL$_N$
black hole entropy turns out to be an additional lift construction
that interprets the Maa\ss-Skoruppa roots $f_{w+2}$ in
terms of modular forms of weight two for arbitrary $N$. It can be shown that the set of 
Maa\ss-Skoruppa roots of the  CHL$_N$ models decomposes into two 
distinct classes of forms, one class admitting
complex multiplication, the other not. For this reason it
is clear that the lifts of weight two modular forms to
the higher weight forms $f_{w+2}$ must involve two different
constructions, depending on the type of the higher weight forms $f_{w+2}$.
For the forms without complex multiplication the lift
interpretation of $f_{w+2}$  in terms of weight
two form $f_2 \in S_2$ can be shown to be given by the following relation 
 \begin{equation}
  f_{w+2}(q) ~=~ f_2(q^{1/m})^m, ~~~~{\rm with}~~m =
  \frac{1}{2}\left\lceil \frac{24}{N+1}\right\rceil \in {\mathbb N},
  \end{equation}
 where $\lceil \cdot \rceil$ is the ceiling function.
 The lift for the class of Maa\ss-Skoruppa roots with complex
 multiplication derives from the existence of algebraic Hecke
 characters $\psi_H$ of weight one. The $L$-functions of powers $\psi_H^{w+1}$ of 
these characters are the inverse Mellin transforms of  the forms $f_{w+2}$ \cite{cs11}.

 The interpretation of all CHL$_N$ Maa\ss-Skoruppa roots $f_{w+2}$ in terms of weight two modular
 forms $f_2^{\widetilde N}$ via  the additional lifts just described 
 implies that the motivic orgin of the Siegel modular entropy of CHL$_N$ models
 is to be found in elliptic curves, as opposed to higher dimensional geometric 
 structures associated to the compactification manifold. 
This follows from the fact that for each of the models it is possible to 
 find an elliptic curve $E_{\widetilde N}$, whose conductor ${\widetilde N}$ 
 depends on the order $N$ 
of the quotient group ${\mathbb Z}_N$, such that the modular form $f(E_{\widetilde N},q)$ 
 associated to $E_{\widetilde N}$   via its $L-$function agrees with the modular form 
$f_2^{\widetilde N}$:
  \begin{equation}
   f_2^{\widetilde N}(q) ~=~ f(E,q). 
  \label{modular-identity}\end{equation}
Abstractly, this follows from the proof of by Wiles et. al. of the
 Taniyama-Shimura-Weil conjecture \cite{w95,tw95,bcdt01},
 but no such heavy machinery is necessary for the concrete cases
 based on the CHL$_N$ models, where the elliptic curves $E_{\widetilde N}$ 
can be constructed explicitly 
  for each order $N$ \cite{cs11}.
 
 The enhanced lift construction 
  \begin{equation}
     f_2^{\widetilde N} ~\longrightarrow ~ f_{w+2}^N  ~\longrightarrow ~ \varphi_w^N  
              ~\longrightarrow ~ \Phi_w^N,
   \end{equation} 
  extending the Maa\ss-Skoruppa lift (\ref{ms-lift}), in combination with the 
 modular identity (\ref{modular-identity}) shows that the motivic origin of the
 Siegel black hole entropy of the class of CHL$_N$ models arises from
 lower-dimensional cycles. This is surprising because one might have expected that 
it is the three-dimensional
 structure of the compactification manifold $X_N=T^6/{\mathbb Z}_N$ in the
 heterotic frame, or $(K3\times T^2)/{\mathbb Z}_N$ in the type IIA
 frame, that supports the black hole entropy of the CHL$_N$ black holes.

\vskip .3truein

 \noindent
{\bf Acknowledgement} \hfill \break
 The work described here was supported in part by the National Science Foundation
 under grant No. PHY 0969875. It is a pleasure to acknowledge further support from  
 the Max Planck Werner Heisenberg Institute in Munich, as well as from CERN, and 
 to thank in particular   Dieter L\"ust and Wolfgang Lerche for making these visits possible.
 I'm grateful to the theory groups at both institutions for their friendly hospitality.

\vskip .2truein

\bibliographystyle{ws-rv-van}

\end{document}